\documentclass[fleqn,10pt]{wlscirep}
\usepackage{adjustbox}
\usepackage[utf8]{inputenc} % allow utf-8 input
\usepackage[T1]{fontenc}    % use 8-bit T1 fonts
\usepackage{hyperref}       % hyperlinks
\usepackage{url}            % simple URL typesetting
\usepackage{booktabs}       % professional-quality tables
\usepackage{amsfonts}       % blackboard math symbols
\usepackage{nicefrac}       % copact symbols for 1/2, etc.
\usepackage{microtype}      % microtypography
\usepackage{graphicx}
\usepackage{placeins}
\usepackage[singlelinecheck=false]{subcaption}
\usepackage{multirow}
\usepackage{mathtools}
\usepackage{algorithm}
\usepackage{algpseudocode}
\usepackage{siunitx}

\usepackage{xspace}
\usepackage{amssymb}
\newcommand{\molformer}{\textsc{MoLFormer}\xspace}

\definecolor{rc1}{RGB}{4, 142, 255}
\definecolor{rc2}{RGB}{255, 104, 3}

\newcommand{\RNum}[1]{\uppercase\expandafter{\romannumeral #1\relax}}

\title{GP-\molformer: A Foundation Model For Molecular Generation}

\author[1,*]{Jerret Ross}
\author[1]{Brian Belgodere}
\author[1]{Samuel C.~Hoffman}
\author[1]{Vijil Chenthamarakshan}
\author[1]{Jiri Navratil}
\author[1]{Youssef Mroueh}
\author[1,*]{Payel Das}
\affil[1]{IBM Research, Yorktown Heights, NY 10598, USA}

\affil[*]{rossja@us.ibm.com and daspa@us.ibm.com}

\begin{abstract}

Transformer-based models trained on large and general purpose datasets consisting of molecular strings have recently emerged as a powerful tool for successfully modeling various structure-property relations. Inspired by this success, we extend the paradigm of training chemical language transformers on large-scale chemical datasets to generative tasks in this work. Specifically, we propose GP-\molformer, an autoregressive molecular string generator that is trained on more than 1.1B (billion) chemical SMILES. GP-\molformer uses a 46.8M parameter transformer decoder model with linear attention and rotary positional encodings as the base architecture.   GP-\molformer's utility is evaluated and compared with that of existing baselines on three different tasks: \textit{de novo} generation, scaffold-constrained molecular decoration, and unconstrained property-guided optimization. While the first two are handled with no additional training, we propose a parameter-efficient fine-tuning method for the last task, which uses property-ordered molecular pairs as input. We call this new approach \emph{pair-tuning}. Our results show GP-\molformer performs better or comparable with baselines across all three tasks, demonstrating its general utility for a variety of molecular generation tasks. We further report strong memorization of training data in GP-\molformer generations, which has so far remained unexplored for chemical language models. Our analyses reveal that training data memorization and novelty in generations are impacted by the quality and scale of the training data; duplication bias in training data can enhance memorization at the cost of lowering novelty. We further establish a scaling law relating inference compute and novelty in generations.
\end{abstract}

\begin{document}
\maketitle

\section*{Main}\label{sec:introduction}

Identifying molecules with desirable properties from the vast landscape of possibilities is daunting. As the search space is enormous and high-throughput screening of molecules is costly and time-consuming, this requires a thorough understanding of the chemical data manifold.
Recently, similar to what has been experienced in computer vision and natural language processing, deep generative models have made great strides in modeling molecular distributions and sampling new molecules from them in \textit{de novo} or targeted manners. Among those efforts, a significant fraction use string-based representations of molecules as the input; thus, techniques explored in language modeling, such as causal and masked language modeling, are becoming widely used in building molecular deep neural models.

Interestingly, much of the recent performance gains for natural language models have come from training at scale --- in terms of the number of parameters and the number of training samples \cite{kaplan2020scaling, sorscher2022beyond, ghorbani2021scaling}. It is reported that larger language models that can memorize training data show improved generalization \cite{tirumala2022memorization, carlini2023quantifying, kandpal2023large}. Furthermore, data that is seen during training many times is memorized more and de-duplication of training data plays a big role in preventing such memorization ~\cite{tirumala2022memorization, carlini2023quantifying, lee2022deduplicating, kalai2024calibratedlanguagemodelshallucinate}.

However, less work has been done in understanding the impact of training data  scale and its memorization on the performance of generative models of molecules.  Specifically, it remains under-explored to what extent a causal large language model of molecules, trained on large-scale (>100M) training data, memorizes its training data and demonstrates such memorization in its generations.  In chemical language modeling tasks, molecules in training data originate from publicly available databases such as ZINC~\cite{irwin2005zinc} and PubChem~\cite{kim2016pubchem}. It is known that certain molecules as well as certain molecular features are over-represented in those databases \cite{jia2019anthropogenic} but how such training bias is perpetuated by generative chemical language models remains relatively unknown.

An additional dimension of scaling in traditional large language models that has been investigated recently is inference-time compute scaling \cite{brown2024largelanguagemonkeysscaling}. It has been shown that with increasing inference compute, performance across multiple tasks can be increased for the same model, as it allows better coverage of the search space. On the other hand, the effect of inference scaling by increasing number of generations is under-explored for molecular generative models.

\begin{figure}[t]
    \centering
    \includegraphics[width=\textwidth]{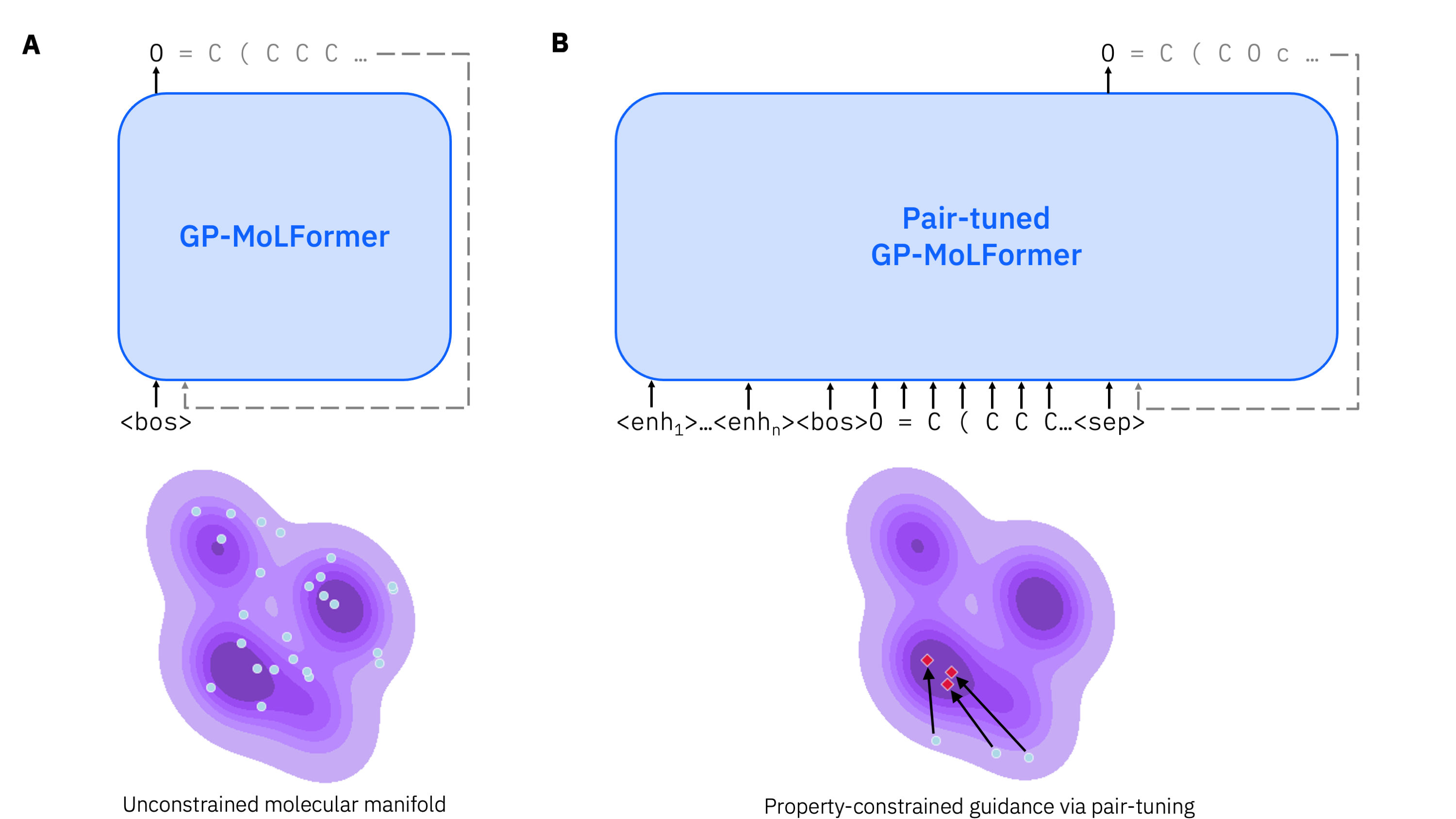}
    \caption{GP-\molformer --- a generative pre-trained molecular foundation model. \textbf{(A)} Unconditional generation using GP-\molformer. SMILES representations are generated autoregressively and randomly along the learned manifold (purple area). \textbf{(B)} During pair-tuning, a prompt vector is learned, which translates a given molecular representation (light blue dots) to an optimized region of the manifold (red diamonds).}
    \label{fig:overview}
\end{figure}

To bridge these gaps, in this work, we present a family of generative pre-trained molecular foundation models for the unconstrained and targeted generation of novel molecules. These decoder-only models are based on the recently published Molecular Language transFormer (\molformer) architecture \cite{ross2022large}. We refer to these Generative Pre-trained models as GP-\molformer. The base transformer architecture of our GP-\molformer consists of $\approx47$M parameters and uses an efficient linear attention mechanism together with rotary positional encodings --- analogous to \molformer\cite{ross2022large} but using decoder instead of encoder blocks (Figure~\ref{fig:overview}A). The model is then trained with a causal language modeling objective on a large corpus of 0.65--1.1 billion canonicalized SMILES strings of small molecules from publicly available chemical databases.

We evaluate  GP-\molformer on an unconditional \textit{de novo}  generation task as well as to two targeted molecular design tasks: scaffold-constrained molecular decoration and unconstrained property-guided optimization. For scaffold decoration, we exploit GP-\molformer's causal language modeling ability and establish GP-\molformer's ability to handle the task without undergoing any task-specific tuning. For the optimization task, we provide a prompt-tuning or soft prompt-learning algorithm that learns from  partial orderings of molecules. We name this method \emph{pair-tuning} (Figure~\ref{fig:overview}B). Results show that pair-tuning on GP-\molformer provides on par or better performance in three different property optimization tasks, namely (i) drug-likeness optimization, (ii) penalized logP optimization, and (iii) optimization of dopamine type 2 receptor binding activity.

We further extensively evaluate quality of GP-\molformer -generated molecules, in the light of the training data scale and the bias present in the training data. Experiments reveal significant memorization in \textit{de novo} generations affecting novelty therein. We further analyze how representational bias encoded in the public chemical databases is perpetuated by a generative chemical language model and is reflected in its generation quality. To our knowledge, this is the first report on effect of training data memorization in a generative pre-trained chemical language model. Further, we investigate the effect of inference compute as another scaling dimension by increasing the number of generated samples and establish a inference scaling law relating number of generations with novelty in them.  Experiments demonstrate that novelty in \textit{de novo} generations by GP-\molformer drops when number of generated samples reaches a scale of $\approx$ 1B. Nevertheless, GP-\molformer is able to generate novel, unique, diverse, and valid molecules even when the generation pool reaches a size of 10B, while showing consistent memorization of training data.

Our main contributions are:
\begin{itemize}\setlength{\itemsep}{-0.6pt}
\item We present an autoregressive transformer-based SMILES decoder, GP-\molformer, trained on up to 1.1 billion chemical SMILES.
\item We report the performance of GP-\molformer models on \textit{de novo} and scaffold-constrained generation tasks.
\item  We provide a parameter-efficient finetuning method, which utilizes property-ranked molecule pairs as input, for property-guided molecule generation and show its effectiveness on three different tasks.
\item We further study training and inference scaling of GP-\molformer and report the effect on generation novelty. We reveal that training data duplication significantly affects novelty in generations. We also report a scaling behavior relating inference compute and novelty that follows exponential decay, while showing that GP-\molformer can generate a notable fraction of novel SMILES, even when number of generations reaches 10B.
\item We demonstrate higher novelty and diversity in GP-\molformer generations compared to baselines, which is attributed to its training data scale.
\end{itemize}

\section*{Results and Discussion}\label{sec:experiments}

GP-\molformer uses a causal modeling objective of predicting the next token given the context history of prior tokens in the input SMILES strings. For details of model architecture and training, see the Methods section. After pre-training, we quantitatively assess the performance of GP-\molformer on  \textit{de novo} molecule generation and scaffold-constrained molecule decoration tasks,  before applying a novel prompt-tuning algorithm for molecular optimization.

\subsection*{\textit{De novo} generation of molecules}

The task under consideration is to generate random, valid molecules by sampling from the generative model. As downstream optimization may rely on these randomly generated molecules as starting points, the generated distribution must contain novel, diverse, and unique molecules. Here we compare a GP-MoLFormer model trained on 650M \emph{unique} SMILES, that is a de-duplicated subset of the 1.1B SMILES extracted from ZINC and PubChem, as described in Ross et al.~(2022)\cite{ross2022large}. This model variant is referred to as GP-\molformer-\textsc{Uniq} hereafter.

We compare GP-\molformer-\textsc{Uniq} with different baseline models, such as the character-level recurrent neural network (CharRNN) \cite{polykovskiy2018molecular}, SMILES variational autoencoder (VAE) \cite{polykovskiy2018molecular},  junction tree VAE (JT-VAE) \cite{jin2018junction}, latent inceptionism on molecules (LIMO) \cite{eckmann2022limo}, and MolGen-7b \cite{fang2024domainagnostic}. Except MolGen-7b, all baselines were trained and tested on datasets from MOSES\cite{polykovskiy2018molecular}, whose origin is the ZINC Clean Leads dataset\cite{irwin2005zinc}. The size of that training set is 1.6M. MolGen-7b was trained on 100M filtered molecules from ZINC-15\cite{irwin2005zinc} as detailed in Irwin et al.~(2022)\cite{irwin2022chemformer}. LIMO and MolGen-7b are trained using an alternative molecular string representation, SELFIES \cite{Krenn_2020_selfies}, that guarantees 100\% validity of generated molecules. We, in contrast, train GP-\molformer-\textsc{Uniq} on SMILES as recent work shows that training a generative language model on SELFIES may hurt model's exploratory ability \cite{skinnider2024invalid}. All baseline performances are reported on their corresponding test set consisting of 175k molecules (if the original test set was larger, this is a randomly selected subset).

First, we note that GP-\molformer-\textsc{Uniq} (and GP-\molformer) exhibits excellent validity and uniqueness at standard generation size (30/10k). See Supplementary Table \ref{tab:val_nov_uniq} for comparison with the baseline models. At the same time, we argue that these metrics are insufficient to measure generation \emph{at scale}. Furthermore, as we show later, novelty is dependent on training set size in addition to generation size so models trained on different size datasets are not directly comparable. See the Scaling results section below for further discussion.

Standard metrics for evaluating model-generated molecules are reported in Table \ref{tab:moses} for a generation set of 30k  molecules. When compared to baselines, GP-\molformer-\textsc{Uniq} is equally performant in generating molecules that share high cosine similarity with the corresponding reference molecules at the fragment (Frag) level, consistent with low Fréchet ChemNet Distance (FCD)\cite{preuer2018frechet}. At the same time, GP-\molformer-\textsc{Uniq} generates molecules with high internal diversity (IntDiv), i.e., average pairwise dissimilarity. The scaffold cosine similarity (Scaf) and similarity to the nearest neighbor in the test set (SNN) of GP-\molformer-\textsc{Uniq} is comparable to that of baselines for 30k generations. All these metrics are computed using the MOSES\cite{polykovskiy2018molecular} framework (we limit our scope to MOSES in this study, although we note that myriad other benchmarks are available for evaluating generative molecular models\cite{brown2019guacamol,nigam2023tartarusbenchmarkingplatformrealistic,huang2022artificial}).

We further report analogous metrics computed using \molformer\cite{ross2022large} embeddings as the chemical features  and estimate distances using those embeddings as a measure of similarity (under column \molformer-based metrics; see Table \ref{tab:moses} caption for details). The trends observed on these metrics further support the fact that GP-\molformer-Uniq generates a molecular distribution that is close to the training in terms of fragment and scaffold composition as well as projections to \molformer space, while exhibiting high diversity, when compared to baselines.

\renewcommand{\thefootnote}{\fnsymbol{footnote}}
\begin{table*}[t]
\centering
\begin{tabular}{lrrrrrrrr}
\toprule
& \multicolumn{5}{c}{MOSES metrics} & \multicolumn{3}{c}{\molformer-based metrics} \\
\cmidrule(lr){2-6}\cmidrule(lr){7-9}
&  Frag$\uparrow$ & Scaf $\uparrow$ & SNN$\uparrow$ &  IntDiv$\uparrow$ & FCD$\downarrow$ & DNN$\downarrow$ & IntDiv2$\uparrow$ & FMD$\downarrow$ \\
\midrule
CharRNN \protect{\cite{polykovskiy2018molecular}} & \textbf{0.9998} & 0.9242 & 0.6015 & 0.8562 & 0.0732 & 5.735 & 13.03 & 0.1515 \\
VAE \protect{\cite{polykovskiy2018molecular}} &  0.9984 & \textbf{0.9386} & \textbf{0.6257} & 0.8558 & 0.0990 & \textbf{5.549} & 13.09 & 0.2531 \\
JT-VAE \protect{\cite{jin2018junction}}  &  0.9965 & 0.8964 & 0.5477 & 0.8551 & 0.3954 & 6.312 & 12.97 & 1.700 \\
LIMO \protect{\cite{eckmann2022limo}} & 0.6989 & 0.0079 & 0.2464 & \textbf{0.9039} & 26.78 & 11.41 & 13.08 & 162.0 \\
\midrule
MolGen-7b \protect{\cite{fang2024domainagnostic}} & \textbf{0.9999} & 0.6538 & 0.5138 & 0.8617 & \textbf{0.0435} & 6.788 & 12.58 & \textbf{0.1237} \\
\midrule
GP-\molformer-\textsc{Uniq} & \textbf{0.9998} & 0.7383 & 0.5045 & 0.8655 & 0.0591 & 6.970 & \textbf{13.10} & 0.1844 \\
\bottomrule
\end{tabular}
\caption{Comparison of 30k generations with a held-out test set of size 175k.
``MOSES metrics'' columns refer to the typical set of generation performance metrics (see Polykovskiy et al.\cite{polykovskiy2018molecular} for details) computed with respect to the following sets:
Baseline performances for CharRNN, VAE, and JT-VAE are taken from MOSES\cite{polykovskiy2018molecular}. LIMO is reproduced using their random generation model trained on the MOSES data. These models all use the default MOSES test split for reference. The MolGen-7b baseline uses the pre-trained model from Hugging Face\protect\footnotemark\ with multinomial sampling ($T=1.0$) and is tested on a random 175k subset of the original test data. GP-\molformer-\textsc{Uniq} is tested with respect to a held-out 175k set from its training data.
``\molformer-based metrics'' columns refer to analogous metrics computed using \molformer\cite{ross2022large} embedding distances instead of Tanimoto similarity. DNN is the average Euclidean distance from generated molecules to the nearest molecule from the test set. IntDiv2 is the average pairwise Euclidean distance between generated molecules. FMD is the Fr\'{e}chet distance between the \molformer embedding distributions. \textbf{Bold} values indicate the best model for a given metric.}
\label{tab:moses}
\end{table*}
\footnotetext{\url{https://huggingface.co/zjunlp/MolGen-7b}}

We also calculated the pairwise Tanimoto similarity between novel and unique generations and molecules from the corresponding 175k sample test set using molecular fingerprints as features. We then report both the average similarity per generated molecule and the maximum similarity per generated molecule over the test set. These results are presented in Table \ref{tab:simnov}. GP-\molformer-\textsc{Uniq} is slightly lower than MolGen-7b in both average mean and average maximum similarity, indicating generations are slightly more dissimilar with respect to its test set. LIMO results are much lower than both of these, though, as we see in Table~\ref{tab:moses}, the outputs of this model do not match its test set well, so this is to be expected. We also provide discussion of scaffold novelty in Supplementary Figure \ref{fig:similarity}. Taken together, these analyses mean that GP-\molformer-\textsc{Uniq} is able to generate molecules that are distinct at the scaffold level as well as at the fingerprint level, to a similar extent to other models, despite the seemingly low string level novelty scores in the Supplementary Table~\ref{tab:val_nov_uniq} (see also the discussion of the Supplementary Figure~\ref{fig:samplingtemp}).

\begin{table*}[t]
    \centering
    \begin{tabular}{lrr}
    \toprule
    Model & mean & max \\
    \midrule
    LIMO & 0.0905 & 0.2474 \\
    MolGen-7b & \textbf{0.1385} & \textbf{0.5138} \\
    GP-\molformer-\textsc{Uniq} & 0.1354 & 0.4533 \\
    \bottomrule
    \end{tabular}
    \caption{Average Tanimoto similarity between 30k randomly generated molecules and all respective test molecules (mean) or most similar test molecule (max). \textbf{Bold} values indicate the highest similarity to test molecules.}
    \label{tab:simnov}
\end{table*}

\begin{figure}[th]
    \centering
    \includegraphics[width=0.75\textwidth]{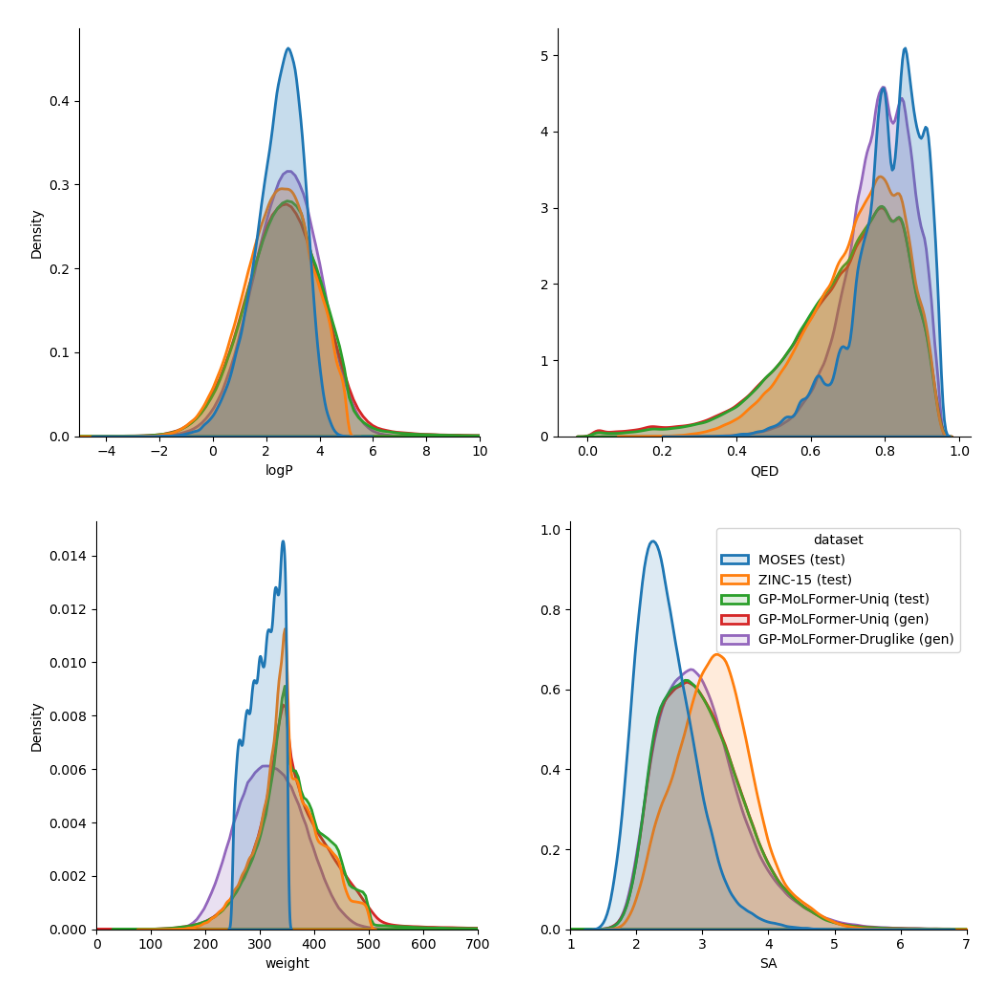}
    \caption{Property distributions of different test datasets --- MOSES, ZINC-15 (MolGen-7b), and GP-\molformer-\textsc{Uniq} (ours) --- along with generated samples from GP-\molformer-\textsc{Uniq} and GP-\molformer-\textsc{Druglike}. Clockwise from top left: octanol-water partition coefficient, drug-likeness, synthetic accessibility, molecular weight. Our test distributions are consistently wider (more diverse) than the other baselines. Furthermore, the generated distribution matches the corresponding test distribution almost exactly. In comparison to GP-\molformer-\textsc{Uniq}, a density shift toward higher QED values with GP-\molformer-\textsc{Druglike} can be observed, as expected.}
    \label{fig:dataset_property_distributions}
\end{figure}

Figure \ref{fig:dataset_property_distributions} shows the property distributions of the different test sets, as well as of molecules generated using  GP-\molformer-\textsc{Uniq}. The generated distribution shows very good reproduction of the corresponding test distribution. Furthermore, while GP-\molformer-\textsc{Uniq}'s performance is estimated on a held-out test set that is of similar size, we found this test set to be more diverse in terms of number of unique scaffolds present within the set (126k compared to 124k in the ZINC-15 subset and 77k in the MOSES set) and by comparing different property distributions with that of the other baselines. More analyses on how these statistics change with training data variations and generated pool size can be found later (see Discussions).

We also examine the domain adaptation of GP-\molformer via down-stream finetuning on a set of 36.7M drug-like molecules from PubChem~\cite{kim2016pubchem}.
In Figure \ref{fig:dataset_property_distributions}, we show results of this fine-tuned model, referred as  GP-\molformer-\textsc{Druglike}. The fine-tuning set contains molecules with QED $>0.6$ \cite{lee_2023_zenodo}. Results show that, the generated molecules undergo a distribution shift in properties as expected. For example, the QED distribution is shifted toward right compared to GP-\molformer-\textsc{Uniq}. We also provide examples of de novo generated molecules in the Supplementary Figure \ref{fig:samples_denovo}.

\subsection*{Scaffold-constrained decoration}
We further subject GP-\molformer to the task of scaffold-constrained generation. For this experiment, we utilize the GP-\molformer trained on 1.1B SMILES. We first take the five unique scaffolds from the dopamine receptor $\mathrm{D}_2$ (DRD2) active binder dataset validation split\cite{arus2020smiles}. These scaffolds contain between two to four attachment points.  We perform a pre-processing step for each unique scaffold, generating every possible randomized SMILES representation of that scaffold. Then we sort the resulting candidates according to the distance of the ``*'' characters, also known as the attachment point,  to the end of the string.  For multiple *s, we sum the distances.  As an example, \texttt{C1(=O)N(CCN1*)*} would score $2 (2+0)$ while \texttt{C1N(*)C(=O)N(*)C1} would score $15 (12+3)$. If multiple representations are equivalently optimal, we save all of them.  During the generation step, we provide the candidates produced in this pre-processing step as input to GP-\molformer.

Next, the task is to generate multiple possible candidates for the first attachment point given an input scaffold. First, we collect all valid candidates from that generation. Then, we again generate multiple possible candidates for the recently extended scaffolds.  This process is repeated until all the attachment points are decorated then we collect all valid molecules generated.

We compare the performance of GP-\molformer in terms of generating DRD2 active molecules that will pass the DRD2 binding classifier ($p>0.5$). For baselines of comparison, we consider our own random generations from GP-\molformer, as well as an earlier scaffold-conditioned generation model \cite{arus2020smiles} that was specifically trained for scaffold decoration tasks and was then used to decorate the same scaffolds under investigation here with fragments from ChEMBL. In contrast to this baseline model, GP-\molformer has not seen scaffold-constrained generation task during pre-training, nor is it specifically fine-tuned for this purpose. Table \ref{tab:scaffold} shows that GP-\molformer generates more DRD2 active hits compared to a random baseline of \textit{de novo} generation, as well as a generative model trained on this specific task. Examples of scaffold-decorated molecules using GP-\molformer are shown  in the Supplementary Figure \ref{fig:samples_scaffold}.

\begin{table*}[t]
\centering
\begin{tabular}{lc}
\toprule
 & predicted active hits (\%) \\
\midrule
Scaffold decorator\cite{arus2020smiles} & 3.64 \\
\midrule
\textit{de novo} GP-\molformer & 0.83 \\
scaffold-conditioned GP-\molformer & \textbf{4.58}\\
\bottomrule
\end{tabular}
\caption{Scaffold-constrained generation. Predicted active hits is the percentage of generated molecules that pass the DRD2 binding classifier. Baseline performance is taken from Ar\'us-Pous et al.~(2020)\cite{arus2020smiles}. \textbf{Bold} value indicates the best performing model.}
\label{tab:scaffold}
\end{table*}

\subsection*{Unconstrained single property optimization}
Given GP-\molformer demonstrates desirable performance in both novel molecule generation and scaffold-constrained molecular decoration, it makes sense to extend GP-\molformer to downstream task settings, where the goal is to generate molecules with a desired property. In light of current LLM adaptation efforts, one obvious path is model tuning (or ``fine-tuning''), where all model parameters are tuned during adaptation. This approach often is highly data-hungry. As an alternative, prompt-tuning or ``soft prompt'' learning has been proposed, which includes an additional $n$ tuneable tokens for each downstream task, which is prepended to the input text \cite{lester2021power}. This soft prompt is then trained end-to-end on a labeled dataset, whereas the pre-trained LLM remains frozen. This method has been demonstrated to close the gap in model tuning, even when combined with a smaller LLM \cite{lester2021power} and is lower cost compared to full fine-tuning.

We exploit prompt-tuning to introduce a novel means for enabling  GP-\molformer to tackle property-specific molecular optimization tasks, where the goal is to generate molecules with a specific property value above a pre-defined threshold. Below, we describe the pair-tuning framework and then show that pair-tuning performs well on a set of three tasks. We evaluate pair-tuning using GP-\molformer on three property optimization  benchmarks, namely drug-likeness (QED) maximization, penalized logP maximization, and activity maximization for DRD2. The first two properties, QED and penalized logP, are important considerations in drug discovery, and these task shows the ability of a model to optimize salient aspects of a molecule, even if maximization of these properties by themselves is of low utility\cite{zhou2019optimization}. The goal of the third task is to increase the binding affinity of a compound to a ligand, in this case, the dopamine $\mathrm{D}_2$ receptor.

\paragraph{Pair-tuning framework} Following a ``text-to-text'' approach, we formalize the task of generating a (property-)optimal molecule as follows: Given a molecule $a$, translate it to another molecule $b$ with a more optimal property value where $a,b$ come from domain $\Omega$. This conditional generation task is $P_{\theta}(b|a)$, where $\theta$ is the parametrization of the generative language model. This task is handled via learning soft prompts, i.e., prompt-tuning, which is a parameter-efficient task adaptation method for a frozen language model \cite{lester2021power}. Specifically,   we add a small number of task-specific parameters $\phi_{T}$, such that the conditional task becomes $P_{\theta}(b|\phi_{T},a)$ and is trained through maximizing the probability likelihood of $b$. Only $\phi_{T}$ is updated during gradient backpropagation. This procedure is explained in Algorithm~\ref{alg:pair-tuning}.

\begin{algorithm}[t]
    \caption{Pair-tuning training}
    \label{alg:pair-tuning}
    \begin{algorithmic}[1]
        \Require Pre-trained  GP-\molformer; Number of fine-tuning epochs, $m$; number of enhancement tokens, $n$; Pair-Tuning dataset $D=\{(a,b) | b>a\}$ for $a,b\in\Omega$ where $>$ is an order relation defined for a certain property $T$
        \State Append $n+1$ new tokens $\phi_T=\langle enh_1\rangle\langle enh_2\rangle\dots\langle enh_n\rangle,\langle sep\rangle$ to GP-\molformer vocabulary
        \For{$m$  epochs}
            \State Prepare training prompts as $\langle enh_1\rangle\langle enh_2\rangle\dots\langle enh_n\rangle\langle bos\rangle\langle a_1\rangle\langle a_2\rangle\dots\langle a_i\rangle\langle sep\rangle$ where $a=\langle a_1\rangle\langle a_2\rangle\dots\langle a_i\rangle$, i.e., tokenized training molecules
            \State Compute the  cross-entropy (CE) loss conditioned on enhancement tokens $\phi_T$ and molecule $a$ with  the auto-regressive CE loss with target  the molecule $b=\langle b_1\rangle\langle b_2\rangle\dots\langle b_n\rangle \langle eos\rangle$
            \State Compute the gradient of the auto-regressive CE loss with respect to  enhancement tokens $\phi_T$
            \State Update enhancement tokens via gradient descent optimizer
        \EndFor
    \end{algorithmic}
\end{algorithm}

In this formulation, we do not need absolute property values of the molecules, rather only ordered pairs of molecules  are needed. This is to mimic the scenario of many drug and material development tasks, in which two molecules are compared with each other to guide molecular optimization and prioritization, especially for tasks with limited available data. For example, Matched Molecular Pair (MMP) analysis allows the rapid estimation of property differences \cite{dossetter2013matched,yang2023matched}. However, MMP analysis is limited to comparing close molecular derivatives and common molecular derivations, and it can fail to model important chemical contexts. The present formulation of optimizing molecules is free from such constraints and only aims to learn task-specific soft prompts to generate more optimal molecules given a seed molecule.

\begin{table*}[t]
\centering
\begin{tabular}{ccccccccc}
\toprule
& \multicolumn{4}{c}{ Penalized $\log \mathrm{P}$} & \multicolumn{4}{c}{ QED } \\
\cmidrule(lr){2-5}\cmidrule(lr){6-9}
& 1 st & 2nd & 3rd & Validity  & $1 \mathrm{st}$ & 2nd & 3rd & Validity  \\
\midrule
JT-VAE  & 5.30 & 4.93 & 4.49 & 100\% & 0.925 & 0.911 & 0.910 & 100\% \\
MARS  & \textbf{45.0} & \textbf{44.3} & \textbf{43.8} & 100\% & \textbf{0.948} & \textbf{0.948} & \textbf{0.948} & 100\% \\
GRAPHDF  & 13.7 & 13.2 & 13.2 & 100\% & \textbf{0.948} & \textbf{0.948} & \textbf{0.948} & 100\% \\
LIMO on $z$  & 6.52 & 6.38 & 5.59 & 100\% & 0.910 & 0.909 & 0.892 & 100\% \\
LIMO  & 10.5 & 9.69 & 9.60 & 100\% & 0.947 & 0.946 & 0.945 & 100\% \\
$\mathrm{GCPN}$ & 7.98 & 7.85 & 7.80 & 100\% & \textbf{0.948} & 0.947 & 0.946 & 100\% \\
MolDQN-bootstrap & \textbf{11.84} & \textbf{11.84} & \textbf{11.82} & 100\% & \textbf{0.948} & 0.944 & 0.943 & $100\%$ \\
\midrule
Pair-tuning ($k=125$)  & 13.18 (7.12)& 12.24 (6.61) & 11.51 (6.40) &     94.7\% & \textbf{0.948} & 0.947 & 0.947 & 94.7\%  \\
Pair-tuning ($k=1000$)  &    19.59 (9.35) &  15.51(8.93) & 15.27 (8.64) & 94.5\% & \textbf{0.948} & \textbf{0.948} & \textbf{0.948} & 94.5\% \\
\bottomrule
\end{tabular}
\caption{Performance on unconstrained penalized logP and QED optimization. Pair-tuning is performed using frozen GP-\molformer. Baseline performances are taken from Zhou et al.~(2019)\cite{zhou2019optimization} and Eckmann et al.~(2022)\cite{eckmann2022limo} and are reported on 100k generations as per LIMO\cite{eckmann2022limo}. For GP-\molformer, we set $k$, the number of targeted generation attempts per molecule, to $125$ --- given a test set of size 800  this results in 100k total generations. Values in parentheses are after post hoc length filtering. \textbf{Bold} values indicate the highest property values found (both length-constrained and unlimited).}
\label{baseline}
\end{table*}

\paragraph{Penalized logP optimization}
Table \ref{baseline} shows results of pair-tuning on GP-\molformer, as well as of the baselines, in terms of the generated molecules with high penalized logP. Penalized logP is calculated as $\log\mathrm{P}-\mathrm{SA}-{\max(\max_\mathrm{rings}(\mathrm{size}) - 6, 0)}$, i.e., logP penalized by SA and maximum ring size, if larger than 6. We report pair-tuning performances as a function of two different $k$, where $k$ is the number of targeted generation attempts per molecule. For $k=125$, using a test set containing 800 molecules gives a total number of generated molecules of 100k, which is the same used for the baselines. The baselines under consideration are JT-VAE \cite{jin2018junction}, GCPN \cite{you2018graph}, MolDQN \cite{zhou2019optimization}, MARS \cite{xie2021mars}, GraphDF \cite{DBLP:conf/icml/LuoYJ21}, and LIMO \cite{eckmann2022limo}. Penalized logP can be artificially inflated simply by generating molecules with increased length, specifically by adding alkyl carbons \cite{xie2021mars, eckmann2022limo}. Many works, e.g.,  GCPN, MolDQN, and LIMO,  avoid this by reporting top property scores given length constraints, e.g., limiting the length up to the maximum molecule size of the ZINC250k dataset \cite{gomez2018automatic}. MARS, on the other hand, does not consider such a length constraint. We also report the top 3 scores for pair-tuning with a length constraint (length < 38), added post generation,  in Table~\ref{baseline} as the value within parentheses. Compared to the strongest length-constrained baselines, pair-tuning generates molecules with comparably high values. When the length constraint is not considered, pair-tuning still generates molecules with higher but reasonable penalized logP values. Note that pair-tuning does not require feedback or evaluation on generations from an additional reward model or a property predictor, nor is the generative model updated during the tuning. We also report top 3 scores for 1M generations ($k=1000$), which requires less than an hour to generate. Although all the baselines produce molecules with 100\% validity due to their methods utilizing SELFIES or graphs, our method's validity is still very high (around 95\%) and overall this is negligible compared to the ease of generating additional molecules. Altogether, Table \ref{baseline} shows that the proposed method can generate molecules with even higher penalized logP values, both with and without a length constraint.

\paragraph{QED optimization}
As with penalized logP, we show results for QED optimization in Table \ref{baseline} (also see Supplementary Figure \ref{fig:samples_qed} for generated molecules)  compared with the same baselines. Again, pair-tuning performances are reported for two different values of $k$, showing comparable performances with respect to baselines. Supplementary Tables \ref{tab:qed_mol_div} and \ref{tab:plogp_mol_div} further demonstrate that pair-tuning with GP-\molformer produces higher scoring molecules that also share high diversity as well as high closeness to training distribution, compared to baselines.

\begin{table*}[t]
\centering
\begin{tabular}{lcc}
\toprule
 & predicted activity score & average seed score \\
\midrule
Mol-CycleGAN & 0.381 & 0.179 \\
Gargoyles & 0.782 & 0.122 \\
Pair-tuning & \textbf{0.844} & 0.007 \\
\bottomrule
\end{tabular}
\caption{Performance on the unconstrained DRD2 activity optimization task with respect to the initial value. Baseline performances are reported from Erikawa et al.~(2023)\cite{erikawa2023gargoyles}. \textbf{Bold} value indicates the best performing model.}
\label{tab:drd2}
\end{table*}

\paragraph{DRD2 activity optimization}
DRD2 activity optimization results are reported in Table \ref{tab:drd2}. Activity scores are calculated using the trained predictor from Olivecrona et al.~(2017)\cite{olivecrona2017molecular}. Average activity scores of the input seed molecules are also shown. Different baseline performances are reported using different test seed molecules, and we are interested in comparing the activity improvement of an experiment with the set of test seed molecules. For pair-tuning, performance reported considers $k=20$ generations per seed molecule and we use the top 1 for each, similar to the baselines, Mol-CycleGAN \cite{maziarka2020mol} --- a graph-based generation method that uses a CycleGAN optimization scheme --- and Gargoyles \cite{erikawa2023gargoyles}, which uses Monte Carlo Tree Search in fragment space to optimize molecular graphs based on an evaluation function. Results show that pair-tuning generates molecules with the highest activity improvement with respect to the seed molecules, when compared to baselines. Examples of generated molecules using pair-tuning on GP-\molformer are provided  in the Supplementary Figure \ref{fig:samples_drd2}.

\subsection*{Scaling results}
\paragraph{Effect of training dataset and generation pool} The generative ability of a chemical language model can be affected by the scale of the training data. Further, bias in training data can also contribute to the quality of generation. To disentangle these factors, we report GP-\molformer's performance on two independently varying dimensions: First, we vary the \textit{size and quality of pre-training data}. For this dimension, we compare GP-\molformer trained on 1.1B SMILES extracted from ZINC and PubChem, and GP-\molformer-\textsc{Uniq} trained on a 650M de-duplicated subset of that 1.1B SMILES set. Secondly, we analyze the effect of varying the number of generated molecules from 30k to 10B molecules, to study the \textit{effect of inference time scaling}.

To summarize, GP-\molformer is trained on a dataset of 650M--1.1B SMILES, which  captures the relative abundance of molecules, as well as the presence of the same molecule in different context, as found in chemical databases and is evaluated on generations up to a scale of billions. This is in contrast to the existing molecular generation benchmarks that report performance metrics for a relatively small 10--30k generations, and to the current generative molecular models that are designed to target a specific distribution of molecules, e.g., synthetic molecules with biological activity or natural products, and are trained on 1--100M samples\cite{polykovskiy2018molecular}.

We report in Table \ref{tab:comparison_training} the percentage of novel (unseen in training), valid (syntactically correct), and unique (not previously generated) molecules for both GP-\molformer and GP-\molformer-\textsc{Uniq}, for generation size of 30k to 10B. The results show that the fraction of novel generations stays at a consistent $\approx32\%$ for GP-\molformer when the number of total generated molecules is below 1B. Novelty in GP-\molformer-\textsc{Uniq} is $\approx5$--$8\%$ higher compared to that of GP-\molformer for all generation pool sizes. At or beyond 1B generations, the fraction of novel and unique generations drops but still remains significant. Even for 10B generations, GP-\molformer is able to generate a significant 16.7\% novel molecules while GP-\molformer-\textsc{Uniq} is able to generate 21.4\% novel molecule. GP-\molformer, irrespective of training data, outputs chemically valid SMILES almost all the time. While the percentage of valid molecules drops slightly with increasing generation pool size, it still is over 99\% for 10B generations.

Additionally, when comparing the 10B molecules generated by the GP-\molformer and by the GP-\molformer-\textsc{Uniq} model, 67 to 74\% of the novel molecules generated by a model  are unique to that model (i.e., not in the other model's generated set). This implies that the two models learned separate but overlapping manifolds. This aspect of different coverage of the molecular manifold with different model variants will be investigated further in future work.

This result confirms that (i) GP-\molformer trained on a billion of SMILES memorizes training samples, as seen from the high number of exact matches ($1-$novelty, which can be up to 60\%) with training molecules; and (ii) training memorization becomes less when the training data is de-duplicated, enabling more novel generation. (iii) With scaling of inference compute, novelty in generations reduces, but remains significant, even when evaluated against $\approx$10B generations. In summary, in all cases studied here, GP-\molformer is capable of generating novel, diverse, and valid molecules.

\begin{table}[t]
\centering
\begin{tabular}{c c c c c c c }
\hline
 & \multicolumn{3}{c}{Training Size = 650M} & \multicolumn{3}{c}{Training Size = 1.1B} \\
\cmidrule(lr){2-4}\cmidrule(lr){5-7}
Generation Size & Novel & Unique & Valid & Novel & Unique & Valid \\
\hline
30k & 0.390     &      0.997            & 1.000 & 0.323 & 0.997 & 0.997 \\
100k & 0.393 &     0.996 &          0.999 & 0.326 & 0.998 & 0.998 \\
1M  & 0.395     &   0.996    &        0.999 & 0.323 & 0.996 & 0.997 \\
10M & 0.400       &  0.991 &           0.996 & 0.322 &  0.989 & 0.997\\
100M &   0.385 & 0.947   &     0.996      & 0.327 &  0.989& 0.997    \\
1B   &   0.340   & 0.675  & 0.996  & 0.278 & 0.611  & 0.997 \\
10B &    0.214  & 0.270   & 0.996 &  0.167    &0.223 & 0.997      \\
\hline
\end{tabular}
\caption{Novelty, validity, and uniqueness of different number of generations for GP-\molformer trained with 650M (\textsc{Uniq}) and 1.1B (\textsc{Raw}) training set.}
\label{tab:comparison_training}
\end{table}

\paragraph{Discussions on the effect of training data bias on generations}  Within the 1.1B training set, a notable 45\% of SMILES strings were found duplicates.
Some of this is due to molecules that appear multiple times as different isomers, which are pre-processed into the same SMILES representation in our canonicalization pipeline. There is also some overlap between the two databases --- ZINC and PubChem. This popularity of certain molecules may or may not be related to true factors, such as the molecule being useful in a multitude of applications because of its synthetic ease or lower cost, or a combination of all of those. It is reported that such skewed distribution can also originate from anthropogenic attention bias \cite{jia2019anthropogenic}, where some molecules are studied extensively because scientists themselves or their peers ``like'' them. Nevertheless, such over-represenation of certain molecules, either originating from human cognitive biases, heuristics, social influences, from rationally made choices,  from multiple isomeric instances, or a combination of all of the above, can lead to bias in databases.

Data de-duplication is the first step towards removing such bias, which reduces the concentration of the high density regions of the  data manifold. In this case, de-duplication removes isomeric population information as well as repeated molecules across databases. The de-duplicated data is closer to a  data manifold that has more homogenized density all over.
Training on such a data manifold results in higher novelty in generations, as found for GP-\molformer-\textsc{Uniq} when compared to GP-\molformer, as shown in Table \ref{tab:comparison_training}.

Although many existing molecular generative models trained on a much smaller and much focused datasets have demonstrated near-perfect (100\%) novelty in generations, they are for most part not suitable for studying the trade-off between training data memorization and generation novelty.  Investigating such phenomenon requires studying   a generative chemical (language) model that has been trained on a broader-purpose and much larger dataset at scale. Our experiments attempt to address this under-explored aspect in this study. As shown in Table \ref{tab:comparison_training}, novelty in GP-\molformer generations are lower compared to $\approx100\%$ reported by baselines\cite{polykovskiy2018molecular}, but still sufficiently high for practical use. When compared with recent baselines, GP-\molformer generations are more dissimilar to test molecules (see the earlier sections and Table \ref{tab:moses}), though GP-\molformer's test set is more diverse. And, finally, the low novelty in  GP-\molformer's generations is reflective of modeling its vast training set that represents  the relative usage of molecules in real-world.

Similarly, the present study highlights the importance of studying generated sets of different sizes to obtain a comprehensive view of the quality of generations, particularly when the generative model is trained on data at scale.  As GP-\molformer-\textsc{Uniq} aims to capture a training data manifold of more uniform density, which is enabled by de-duplicating the training SMILES,  we see a 1\% rise in novelty as we increase the number of generated samples from 30k to 10M. A similar observation has been reported in image generation \cite{kynkaanniemi2019improved} and language generation \cite{carlini2021extracting, lee2022deduplicating}. To summarize, novelty in generations is influenced by the support provided by both the training distribution and the generated distribution, and should therefore be assessed relative to the sizes and diversity of those two sets.

These results in Table \ref{tab:comparison_training} complement and support earlier efforts focusing on studying scaling behaviors of chemical language models. One such noteworthy effort along this line is Frey et al.~(2023)\cite{frey2023neural}, where neural-scaling behavior in large chemical models was investigated by studying models with over 1B parameters and a scaling relation following a power law was established between training loss and model parameters. However, the models tested in that work were only pre-trained on datasets of size up to $\approx$ 10M data points, which is very small compared to the size of the chemical universe. In Ross et al.~(2022)\cite{ross2022large}, the scaling behavior of  \molformer, which is a transformer-based molecular encoder built using a masked language modeling objective,  was studied. That work clearly established the scaling behavior underlying adaptation of a pre-trained model across downstream tasks, in which the number of model parameters was up to 47M while the number of training points considered was >1B. It was shown that a \molformer trained on 100M SMILES consistently underperformed across a wide variety of property prediction tasks, including quantum mechanical and physiological, when compared to the model trained on >1.1B SMILES, indicating predictive ability may benefit from such bias in training data. In contrast, the results in Table \ref{tab:comparison_training} show that a generative chemical language model trained on cleaner de-duplicated training data produces more novel generations.

We next investigate how these metrics change with varying  number of generated and test molecules. Table \ref{tab:moses_scale} shows that, with increasing the generated pool size,  scaffold similarity with respect to the test molecules becomes $> 0.9$ while SNN reaches $> 0.5$ when compared against 175k held-out test samples. When a larger test set of 1M molecules is used, further increases in both scaffold similarity and SNN are observed. These results imply that, with increasing size and diversity of the training data, the typical metrics used in assessing molecular generative models, such as various similarity measures with respect to a test set, should be carefully analyzed with  generation and test sets that are larger in size compared to what is typically used in the field. Note that, even for 1M generations, GP-\molformer produces highly diverse molecules.

\begin{table*}[t]
\centering
\begin{tabular}{rrrrrrrrrr}
\toprule
&& \multicolumn{5}{c}{MOSES} & \multicolumn{3}{c}{\molformer} \\
\cmidrule(lr){3-7}\cmidrule(lr){8-10}
Test size & Gen.\ size &  Frag$\uparrow$ & Scaf$\uparrow$ & SNN$\uparrow$ &  IntDiv$\uparrow$ & FCD$\downarrow$ & DNN$\downarrow$ & IntDiv2$\uparrow$ & FMD$\downarrow$ \\
\midrule
\multirow{ 3}{*}{175k} & 30k  & 0.9998 & 0.7383 & 0.5045 & 0.8655 & 0.0591 & 6.970 & 13.10 & 0.1844 \\
& 100k &  0.9998 & 0.8653 & 0.5045 & 0.8657 & 0.0279 & 6.967 & 13.10 & 0.1025 \\
& 1M   &  0.9998 & 0.9375 & 0.5040 & 0.8658 & 0.0178 & 6.970 & 13.11 & 0.0741 \\
\midrule
\multirow{ 3}{*}{1M} & 30k  & 0.9998 & 0.7702 & 0.5738 & 0.8655 & 0.0646 & 6.180 & 13.10 & 0.1684 \\
& 100k  &     0.9998 & 0.9026 & 0.5740 & 0.8657 & 0.0331 & 6.179 & 13.10 & 0.0874 \\
& 1M &     0.9998 &  0.9786 & 0.5739 & 0.8658 & 0.0227 & 6.183 & 13.11 & 0.0600 \\
\bottomrule
\end{tabular}
\caption{Investigation on the interplay between  generation set size and test set size on modeling the molecular distribution of GP-\molformer-\textsc{Uniq} generations. Columns are the same as Table~\ref{tab:moses}.}
\label{tab:moses_scale}
\end{table*}

\paragraph{Scaling law for inference compute}
We further attempt to fit a scaling law to the empirical trend observed in Table~\ref{tab:comparison_training} for novelty with respect to generation size, which reflects the scale of inference compute. For both models, this trend appears to follow an exponential decay of the form:
\begin{equation}
    y = a e^{-bx}
\end{equation}
where $y$ is the novelty, $x$ is the generation size, and $a$ and $b$ are fitted parameters. In practice, $b$ is very close to 0 so we can rewrite this as:
\begin{equation}
    y = a e^{-10^cx}
\end{equation}
for ease of reading. These results can be further seen in Supplementary Figure~\ref{fig:scalinglaw}. We observe that the fitted initial value $a$ is higher and decay constant $b$ is lower for GP-\molformer, meaning it starts with higher novelty and declines more slowly, when compared to GP-\molformer. To our knowledge, this is the first ever investigation of inference scaling of chemical language models.

\section*{Conclusion}
In this work, we investigate the effect of training data bias and diversity on the downstream performance of a generative chemical language model named GP-\molformer, built on top of a recent transformer architecture for chemical language modeling. We show the generality of the proposed GP-\molformer architecture on \textit{de novo} generation and on two distinct targeted design tasks, i.e., scaffold-constrained molecular decoration and unconstrained property-guided molecular optimization, where the model performs better or on par when compared to the baselines. We further show how bias in training data can induce memorization, and thus impact the novelty of generations. We analyze how the commonly used metrics for comparing models' generations with a held-out test set are affected by the diversity of the training distribution; hence, the sizes of the test set and the generation set should be carefully considered before making a conclusion based on those metrics. To our knowledge, this is the first report demonstrating training data memorization and its impact on downstream performance of a generative chemical language model pre-trained on billion-scale data. We also investigate effect of inference compute scaling and establish a scaling law between number of generations and novelty in them.

\section*{Methods}

\subsection*{Model details}\label{model_details}

The GP-\molformer decoder uses the transformer block used in \molformer. To avoid the quadratic complexity associated with regular attention computations in vanilla transformers \cite{vaswani2017attention}, \molformer  utilized   a base 12-layer transformer architecture with linear attention \cite{linear_attention}, wherein each layer has $12$ attention heads and a hidden state size of $768$. A Generalized Random Feature map \cite{FAVOR} for the linear attention was chosen.

To better model positional dependence of tokens within a SMILES string, \molformer deviates from using the default absolute  position embeddings and instead uses rotary embeddings \cite{su2021roformer}: \[ \textnormal{Attention}_{m}(Q,K,V)=\frac{\sum_{n=1}^N \langle  \varphi(R_{m}q_m) ,  \varphi(R_{n}k_n) \rangle v_n }{\sum_{n=1}^N  \langle \varphi(R_{m}q_m) , \varphi(R_{n}k_n) \rangle  },\] where $Q,K,V$ are the query, key, and value respectively, and $\varphi$ a random feature map. The GP-\molformer is trained on the next token prediction task using a cross-entropy objective: $L_{LM}(w1,...,wn)=\sum_{i}{\log{P}(w_{i}|w_{j<i})}$. Given the size of the transformer model and the efficient linear attention, GP-\molformer takes only around 3 milliseconds for a single forward pass during generation, using a single A100 GPU.

\subsection*{Datasets and tokenization}
We used two datasets for pre-training by combining SMILES from the  PubChem \cite{pubchem_kim} and the ZINC \cite{irwin2005zinc} databases with varying proportions from each. The  dataset used in GP-\molformer training contains a total of 1.1B SMILES strings; 111M of them are from the PubChem dataset, whereas the larger 1B portion comes from the ZINC database. The \textsc{Uniq} dataset is a de-duplicated version of that 1.1B size dataset, which comprises 650M SMILES. We utilized the tokenizer from Schwaller et al.~(2019)\cite{SchwallerFWD} to construct a vocabulary. All SMILES sequences from both PubChem and ZINC are converted to a canonical format with no isomeric information using RDKit \cite{rdkit} followed by de-duplication (\textsc{Uniq} only) and tokenization.  All unique tokens extracted from the resulting output give us a vocabulary of $2357$ tokens plus $5$ special tokens, resulting in a total of  $2362$ vocabulary tokens which are used for all pre-trained models considered in this paper, irrespective of pre-training dataset size. The post tokenization sequence length of the molecules range from $1$ to just over $2000$ tokens. We decide to restrict the sequence length range from $1$ token to $202$ tokens, special tokens inclusive, to reduce computation time. Since over $99.4$ percent of all molecules from our dataset contain less than $202$ tokens, we hypothesize that the removal of molecules with more than $202$ tokens would be of minimal negative impact on pre-training.

\subsection*{Large-scale training and parallelization}\label{sec:training}
For pre-training, we use the causal language model objective defined in Devlin et al.~(2019)\cite{devlin-etal-2019-bert}. The training was performed for 4 epochs (for both 1.1B and 650M training dataset sizes) with a fixed learning rate of $1.6\times10^{-4}$ and a batch size of $1600$ molecules per GPU on a total of $16$ A100 80 GB GPUs over $2$ servers connected via EDR Infiniband fabric. It should be noted that as the number of GPUs utilized increased, we found an increase in learning rate was necessary up to a factor of $8$. The GP-\molformer model was trained for 28.75 hours per epoch,  for 115 hours total training time, while the GP-\molformer-\textsc{Uniq} model was trained for 19.75 hours per epoch, less than 80 hours total training time.

In order to scale our training to large datasets (>1B data points), we relied on adaptive bucketing of mini-batches by sequence length, as well as parallelization via distributed data-parallel training. The combination of linear attention, bucketing, and data parallelism allowed us to reduce the number of GPUs needed from roughly 1000 for quadratic attention with no bucketing to 16.

\subsection*{ Pair-tuning}\label{sec:pairtuning}
The paired molecule datasets for pair-tuning experiments were taken from Jin et al.~(2019)\cite{jin2019learning}; The QED paired data used for training consists of 70644 molecule pairs where the first/seed molecule has a QED value in the range of 0.7--0.8 while the second/target molecule has a QED of 0.9--1.0. The penalized logP paired data consists of 60227 molecule pairs. It should be noted that while the paired datasets were collected such that molecular similarity within the pair is 0.4 and 0.6 for QED and logP, respectively, we demonstrate pair-tuning only on unconstrained property optimization tasks --- we do not account for similarity preservation. The test set size for both QED and penalized logP optimization was 800. For the DRD2 binding optimization task, we used 34,404 molecule pairs from ZINC and Olivecrona et al.~(2017)\cite{olivecrona2017molecular} for training and a test set of 1000 molecules\cite{jin2019learning}. For scoring the generated molecules, the bioactivity prediction model from Olivecrona et al.~(2017)\cite{olivecrona2017molecular} is used; inactive compounds were defined with $p < 0.05$ and actives were with $p > 0.5$.

The vocabulary includes 20 randomly initialized prompt embeddings as well as the \texttt{<unk>} embedding from GP-\molformer training. For training, we prepended all 20 prompt embeddings to the \texttt{<bos>} embedding, followed by the embeddings of the first/seed molecule in a specific pair. We then add the \texttt{<unk>} embedding at the end of the first/seed molecule. After the \texttt{<unk>} embedding, we add the embeddings of the target molecule, followed by the \texttt{<eos>} embedding.

For evaluation, we do a forward pass using the following sequence: the first 20 prompt embeddings + the \texttt{<bos>} embedding + the input molecule embeddings + the \texttt{<unk>} embedding. After that, we sample from the token distribution generated by GP-\molformer until \texttt{<eos>} is encountered. For all pair-tuning experiments, batch size was set to 35, the learning rate was fixed at $3\times 10^{-2}$, and the number of epochs run was 1000. Each epoch took
 6 minutes to complete on a single GPU.

\section*{Author contributions}
J.R. and B.B. trained the GP-\molformer models.  S.H. and J.R. performed pair-tuning experiments. S.H., V.C, J.R.,  and  J.N. conducted the model evaluation experiments. P.D., Y.M., J.R., B.B., V.C, and S.H. contributed to the idea generation, study design, result analyses, and paper writing.

\section*{Conflicts of interest}
There are no conflicts to declare.

\section*{Acknowledgments}
Authors acknowledge IBM Research for their support.

\bibliography{confs}
\newpage
\appendix
\renewcommand\thefigure{S\arabic{figure}}
\renewcommand\thetable{S\arabic{table}}
\setcounter{figure}{0}
\setcounter{table}{0}
\onecolumn
\begin{center}
   \LARGE{\textbf{\sffamily Supplementary Information}}
   \end{center}

\section*{Related work}
Earlier work on \textit{de novo} molecule generation mapped a SMILES representation of a molecule into a continuous latent space using character based recurrent neural networks~\cite{polykovskiy2018molecular} and variational autoencoders (VAE)\cite{gomez2018automatic}. The VAE models have an encoder that converts the discrete molecular representation of a molecule into a continuous representation and a decoder that converts the continuous representation back to a discrete representation. The continuous representation allows for efficient \textit{de novo} generation of molecules by sampling from the latent space. A property predictor is often trained on the latent embeddings from the encoder and is used to search for compounds that have specific desirable properties (conditional generation) using gradient-based or Bayesian optimization techniques.

However, there are several drawbacks for this approach. One drawback is that the decoded SMILES string may not always represent a valid molecule, often violating chemical constraints like valency. To alleviate this problem, several approaches have been proposed to augment the model with rules or mechanisms to ensure that the decoded SMILES strings are always valid~\cite{jin2018junction, kusner2017grammar, dai2018syntax, schoenmaker2023uncorrupt}. Another solution is to use alternate more robust string representations like DeepSMILES~\cite{o2018deepsmiles} SELFIES\cite{Krenn_2020_selfies} and Group SELFIES~\cite{cheng2023groupselfies}. While SELFIES-based representation ensures that all the decoded molecules are structurally valid, there is recent work that indicates that invalid SMILES are low-likelihood samples and are beneficial for chemical language models to better model the input molecule distribution~\cite{skinnider2024invalid}.

Another drawback of this approach is that the property predictors trained on the latent embeddings are not as accurate as predictors trained end-to-end on the molecule itself. This necessitates a post-processing step, where the generated molecules have to be filtered using another, more accurate, predictor~\cite{chenthamarakshan2020cogmol}. LIMO~\cite{eckmann2022limo} avoids this problem by freezing the encoder and using a property predictor trained directly on the discrete representation of a molecule to optimize the latent space using an inceptionism-like approach.

Along with string generation methods, several graph and topological generative models have also been proposed~\cite{DBLP:conf/icml/LuoYJ21, jin2020hierarchical,kuznetsov2021molgrow, ma2018constrained, de2018molgan, xie2021mars, you2018graph, schiff2022augmenting}. Another approach to molecule optimization is to use reinforcement learning to generate or modify a molecular graph~\cite{you2018graph, zhou2019optimization, Zhavoronkov2019natbio, olivecrona2017molecular, DBLP:conf/icml/LuoYJ21}. While these methods have been shown to be successful against a variety of molecular optimization tasks, they often require multiple calls to the property predictor during the optimization process and hence are difficult to use along with  computationally expensive property predictors.

\subsection*{Additional generation metrics}

\begin{table*}[ht]
\centering
\begin{tabular}{lllllllll}
\toprule Metrics & Validity & Unique@10k & Novelty\\
\midrule
CharRNN \cite{polykovskiy2018molecular} & 0.975 & 0.999 & 0.842\\
VAE \cite{polykovskiy2018molecular} & 0.977 & 0.998 & 0.695\\
JT-VAE \protect{\cite{jin2018junction}}  & 1.000 & 1.000 & 0.914\\
LIMO \protect{\cite{eckmann2022limo}} & 1.000 & 0.976 & 1.000\\
\midrule
MolGen-7b \protect{\cite{fang2024domainagnostic}} & 1.000 & 1.000 & 0.934\\
\midrule
GP-\molformer & 0.997 & 0.997 & 0.323\\
GP-\molformer-\textsc{Uniq} & 1.000 & 0.997 & 0.390\\
\bottomrule
\end{tabular}
\caption{Comparison of validity, uniqueness, and novelty of 30k (10k for uniqenuess) generations. Baseline performances are taken from \cite{polykovskiy2018molecular,eckmann2022limo}. MolGen-7b results are computed using the open-source weights with multinomial sampling ($T=1.0$). Novelty is computed with respect to the corresponding training set.}
\label{tab:val_nov_uniq}
\end{table*}

\newpage
\section*{Scaffold novelty in generations}
\begin{figure}[ht]
    \centering \includegraphics[width=0.75\textwidth]{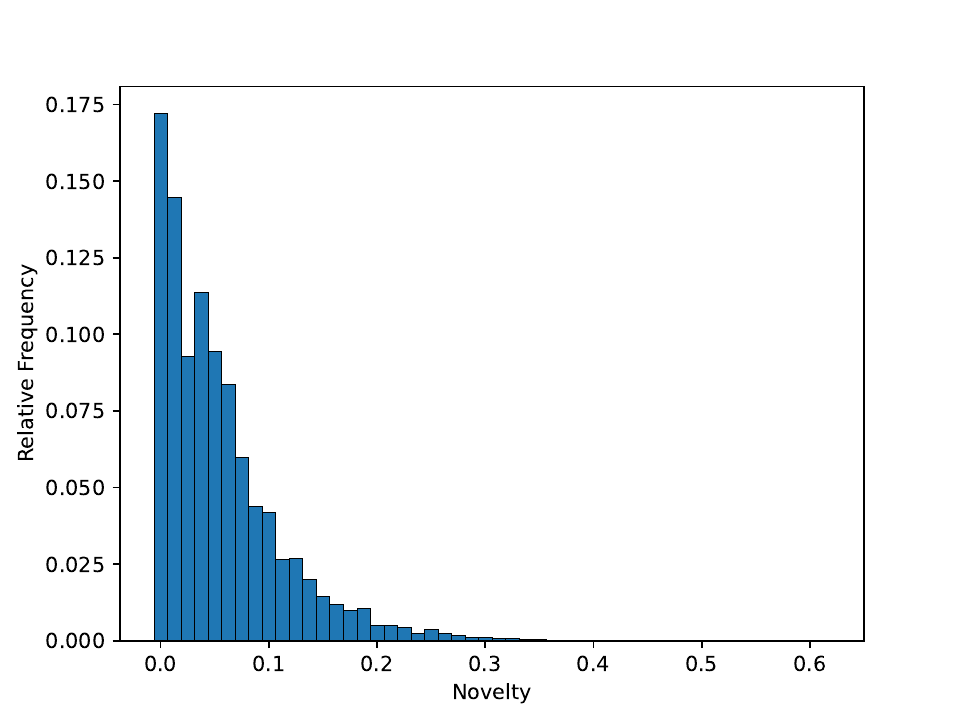}
    \caption{ The novelty of the scaffold of each generated molecule with GP-\molformer-\textsc{Uniq} compared to the most similar scaffold in the training set.}
    \label{fig:similarity}
\end{figure}

In Figure \ref{fig:similarity}, we further compare around 2.2M scaffolds extracted from molecules generated with GP-\molformer with all the scaffolds from the training set (around 1M). While many of the generated scaffolds are present in training set (indicated by a spike at novelty $=0$), we note that there are many novel scaffolds in the generated molecules as well. Even though GP-\molformer is trained on a larger dataset, these results are similar to the results from CogMol\cite{chenthamarakshan2020cogmol}, another generative model trained on a much smaller dataset.

\newpage
\section*{Sampling temperature}

\begin{figure}[!ht]
    \centering
    \includegraphics[width=0.75\textwidth]{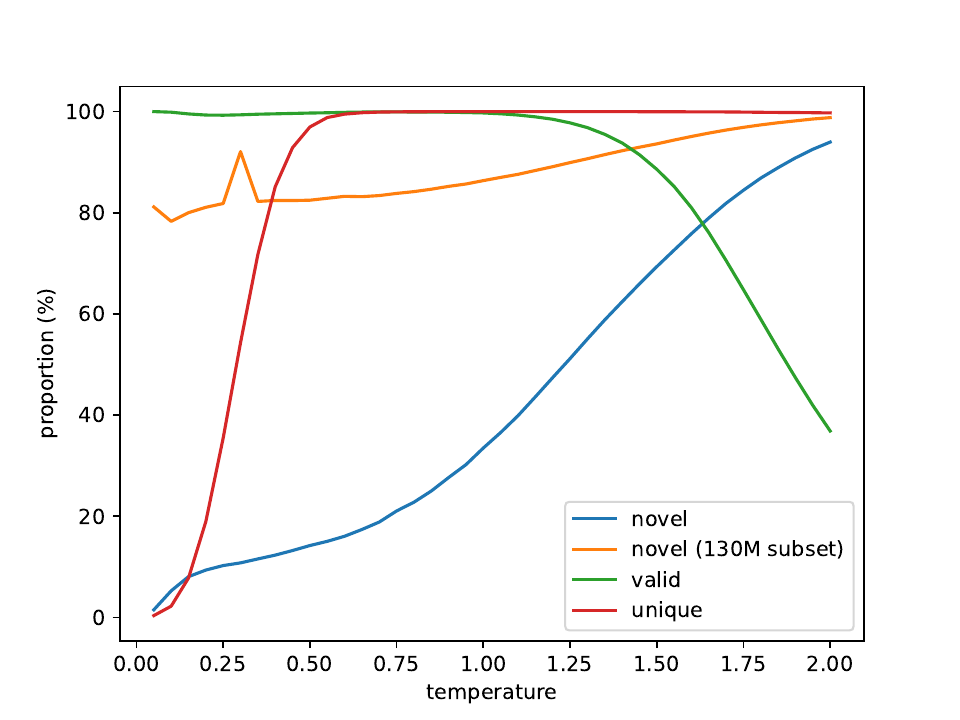}
    \caption{Effect of temperature during multinomial sampling of 100k generated molecules on the percentage of novel molecules (blue) with respect to full 1.1B training set, novelty with respect to a random 130M subset of the training (orange), validity (green), and uniqueness (red) of generations from GP-\molformer.}
    \label{fig:samplingtemp}
\end{figure}

We also report the generative performance of GP-\molformer at several softmax temperature values, as we use multinomial sampling during decoding to encourage novel generation. In fact, recent work has reported the utility of multinomial sampling over beam search for generating molecules that are simultaneously novel and useful (in terms of synthesis and biological testing) \cite{moret2022perplexity}.  Results are shown in Figure \ref{fig:samplingtemp}, which shows that at around $T=1.2$, GP-\molformer performs the best in terms of balancing high novelty (47\%), uniqueness (99\%), and validity (98\%). Further, when we compare novelty of the same generated sets with respect to a random subset of 130M molecules from training set, we see novelty is significantly higher (around 90\% at $T=1.2$), showing the effect of training data scale on generation novelty --- there are simply more opportunities to match with a larger reference set. All results elsewhere in this paper are generated using $T=1.0$.

\newpage
\section*{Samples of \textit{de novo} generated molecules}

\begin{figure}[!ht]
    \centering
    \includegraphics[width=1.0\textwidth]{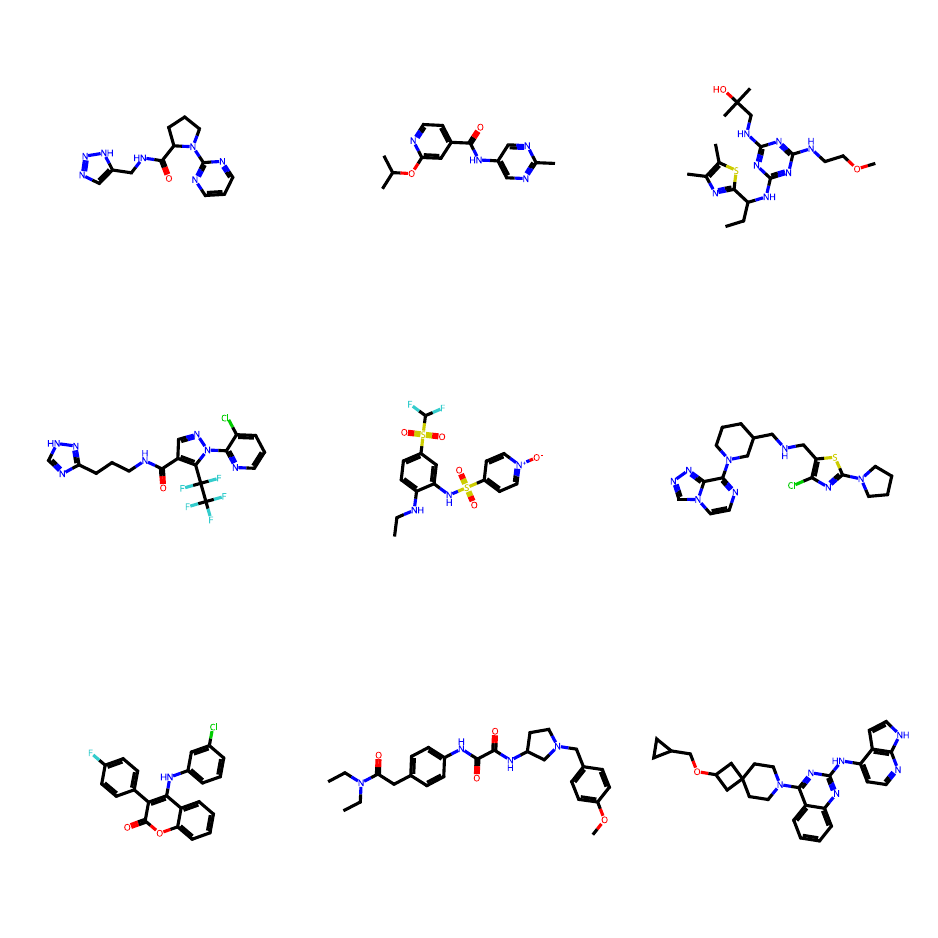}
    \caption{A sample of molecules generated \textit{de novo}.}
    \label{fig:samples_denovo}
\end{figure}

\newpage
\section*{Samples of scaffold-constrained generated molecules}

\begin{figure}[!ht]
    \centering
    \adjustbox{max width=1\textwidth, max height=0.9\textheight, keepaspectratio}
    {
    \includegraphics[width=1\textwidth]{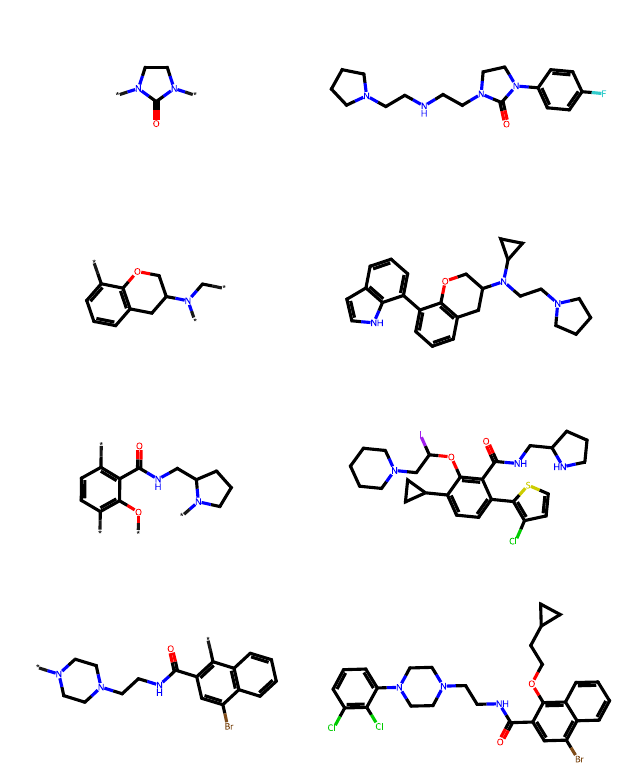}}
    \caption{Scaffold Constrained Generation: Each row shows the scaffold and the generated molecule for that scaffold.}
    \label{fig:samples_scaffold}
\end{figure}

\newpage
\section*{Samples of QED-optimized generated molecules}

\begin{figure}[!ht]
    \centering
    \includegraphics[width=1.0\textwidth]{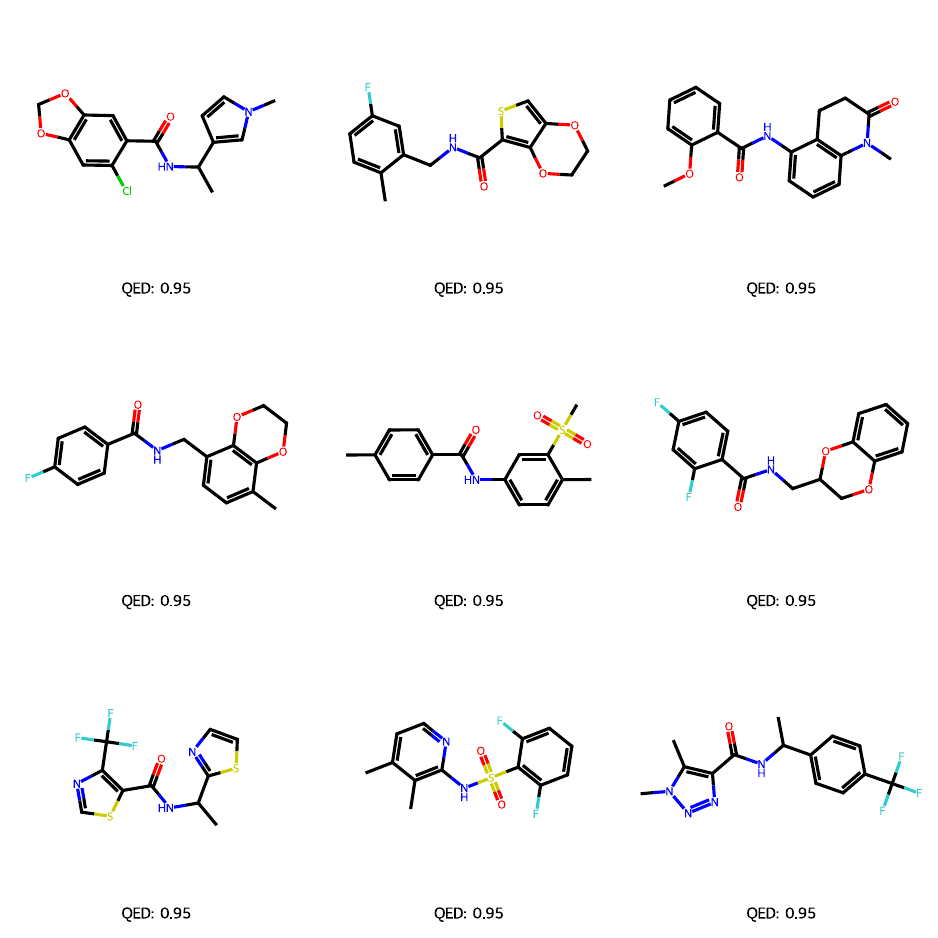}
    \caption{A sample of generated molecules using the pair tuning algorithm for the QED optimization task.}
    \label{fig:samples_qed}
\end{figure}

\newpage
\section*{Diversity analysis in property-optimized molecules}

\begin{table}[h]
\centering
\begin{tabular}{@{}lrrrrrrr@{}}
\toprule
                           & size   & valid & unique & novel & IntDiv & SNN   & best score \\
\midrule
train                      & 88306  & 1.000 & 0.150  & 0.000 & 0.840  & 1.000 & \textbf{0.948}     \\
LIMO                       & 100000 & 1.000 & 0.972  & 1.000 & \textbf{0.906}  & 0.203 & 0.943     \\
Pair-tuning (k=125)        & 100000 & 0.990 & 0.991  & 1.000 & 0.860  & \textbf{0.429} & \textbf{0.948}     \\
Pair-tuning (k=1000)       & 800000 & 0.927 & 0.861  & 1.000 & 0.892  & 0.356 & \textbf{0.948}     \\
\bottomrule
\end{tabular}
\caption{QED molecule diversity. Results for LIMO are reproduced using their published model. Size = \# of examples, valid = fraction of valid SMILES, unique = fraction of unique valid molecules, novel = fraction of novel molecules, IntDiv = internal diversity (p=1) of valid molecules, SNN = average Taniomoto similarity to nearest neighbor in train, best score = maximum QED. \textbf{Bold} values indicate the best models for a given metric.}
\label{tab:qed_mol_div}
\end{table}

\begin{table}[h]
\centering
\begin{tabular}{@{}llrrrrrrr@{}}
\toprule
&                            & size   & valid & unique & novel & IntDiv & SNN   & best score \\
\midrule
\multirow{5}{*}{\rotatebox[origin=c]{90}{unlimited}}
& train                      & 75284  & 1.000 & 0.928  & 0.000 & 0.860  & 1.000 & 5.481     \\
& LIMO                       & 100000 & 1.000 & 0.994  & 1.000 & 0.857  & 0.226 & 6.810     \\
& JT-VAE                     & 2481   & 1.000 & 0.765  & 1.000 & 0.874  & \textbf{0.372} & 5.300     \\
& Pair-tuning ($k=125$)      & 100000  & 0.944 & 0.939  & 1.000 & \textbf{0.896}  & 0.370 & 13.187    \\
& Pair-tuning ($k=1000$)     & 800000 & 0.946 & 0.916  & 1.000 & \textbf{0.896}  & 0.371 & \textbf{19.595}    \\
\midrule
\multirow{5}{*}{\rotatebox[origin=c]{90}{ll-38}}
& train                      & 73154  & 1.000 & 0.926  & 0.000 & 0.860  & 1.000 & 5.288     \\
& LIMO                       & 99618  & 1.000 & 0.994  & 1.000 & 0.857  & 0.226 & 6.768     \\
& JT-VAE                     & 2475   & 1.000 & 0.765  & 1.000 & 0.874  & 0.372 & 4.935     \\
& Pair-tuning ($k=125$)      & 86193  & 1.000 & 0.933  & 1.000 & \textbf{0.898}  & 0.372 & 7.123     \\
& Pair-tuning ($k=1000$)     & 692082 & 1.000 & 0.908  & 1.000 & \textbf{0.898}  & \textbf{0.373} & \textbf{9.355}     \\
\bottomrule
\end{tabular}
\caption{Penalized logP molecule diversity (ll-38 = length-limited to 38 atoms). Results for LIMO are reproduced using their published model while results for JT-VAE use their published set of generated molecules. Size = \# of examples, valid = fraction of valid SMILES, unique = fraction of unique valid molecules, novel = fraction of novel molecules, IntDiv = internal diversity (p=1) of valid molecules, SNN = average Taniomoto similarity to nearest neighbor in train, best score = maximum PLogP. \textbf{Bold} values indicate the best models for a given metric (both unlimited and length-limited).}
\label{tab:plogp_mol_div}
\end{table}

\newpage
\section*{Samples of DRD2-optimized generated molecules}

\begin{figure}[!ht]
    \centering
\includegraphics[width=1\textwidth]{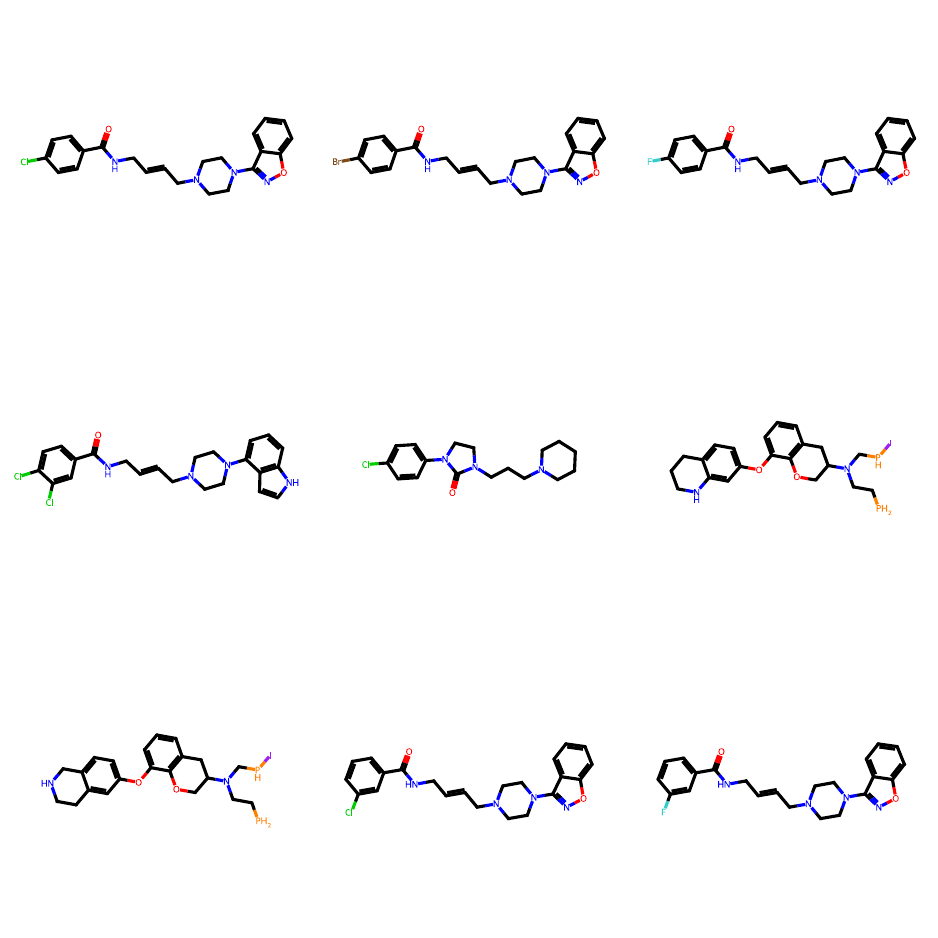}
    \caption{Pair Tuning: Generated Molecules with highest activity to DRD2}
    \label{fig:samples_drd2}
\end{figure}

\newpage
\section*{Scaling law for novelty}

\begin{figure}[h!]
    \centering
    \includegraphics[width=0.75\linewidth]{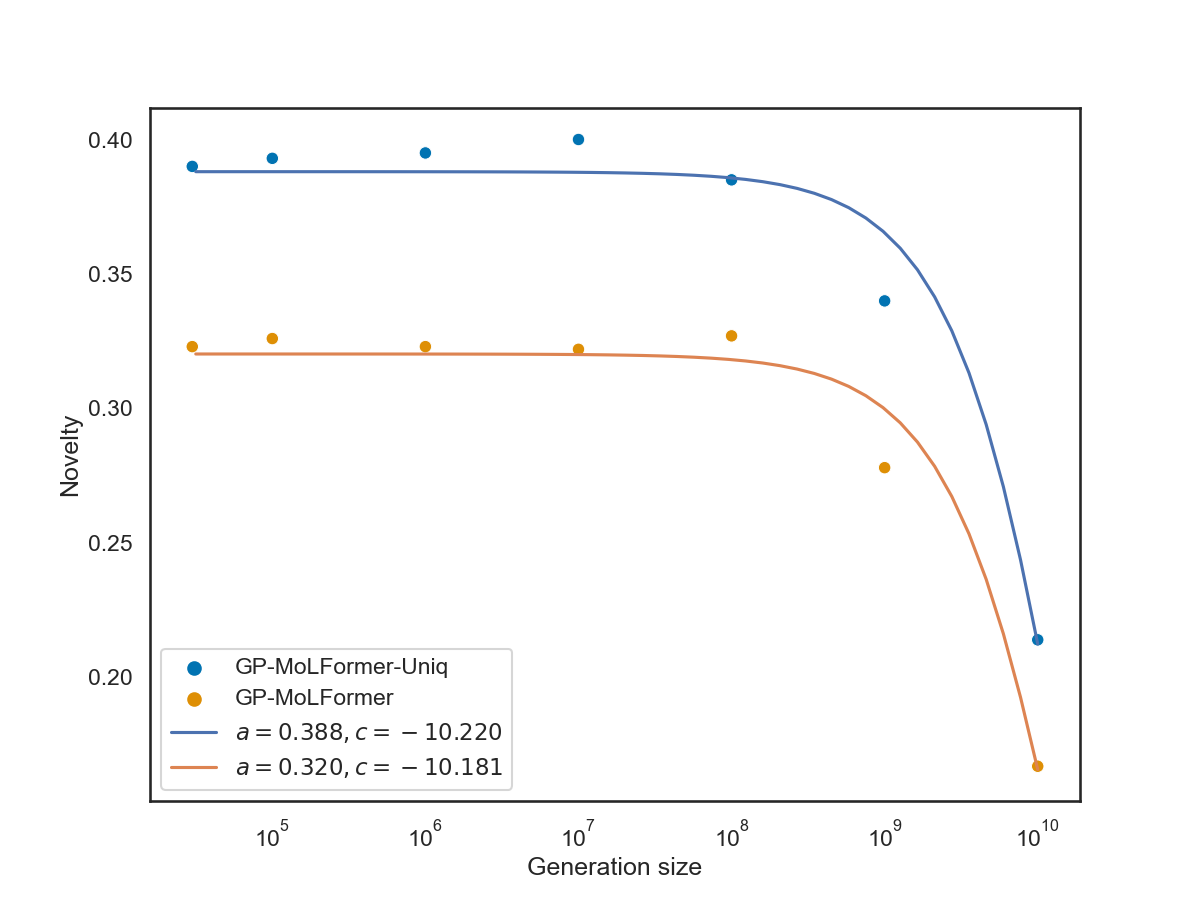}
    \caption{Empirical fit of exponential scaling law on data from Table~\ref{tab:comparison_training}. Besides having a higher initial value ($a$), GP-\molformer-\textsc{Uniq} also has a smaller decay constant ($b=10^c$) meaning it declines in novelty slower as a function of generation size than GP-\molformer.}
    \label{fig:scalinglaw}
\end{figure}

% \clearpage
% \section*{Data Availability Statement}
% We provide a Box folder which contains (1) data and instructions documents, (2) model weights, and (3) code files:\\

% \url{https://ibm.box.com/s/23hyw9hgf13h9lnl3ffo2mgkrnziyaum}\\

% In the document, we provide links to the data used for pre-training as well as the evaluation tasks. We also provide model weights for the pre-trained GP-\molformer and GP-\molformer-\textsc{Uniq}. The accompanying code is also contained in the zip file in the Box folder. Code will be uploaded to GitHub and Zenodo before publication and models will be uploaded to HuggingFace.

\end{document}